# Inter-Patient ECG Classification with Convolutional and Recurrent Neural Networks

Li Guo, Gavin Sim and Bogdan Matuszewski

*Abstract*— The recent advances in ECG sensor devices provide opportunities for user self-managed auto-diagnosis and monitoring services over the internet. This imposes the requirements for generic ECG classification methods that are inter-patient and device independent. In this paper, we present our work on using the densely connected convolutional neural network (DenseNet) and gated recurrent unit network (GRU) for addressing the inter-patient ECG classification problem. A deep learning model architecture is proposed and is evaluated using the MIT-BIH Arrhythmia and Supraventricular Databases. The results obtained show that without applying any complicated data pre-processing or feature engineering methods, both of our models have considerably outperformed the state-of-the-art performance for supraventricular (SVEB) and ventricular (VEB) arrhythmia classifications on the unseen testing dataset (with the F1 score improved from 51.08 to 61.25 for SVEB detection and from 88.59 to 89.75 for VEB detection respectively). As no patient-specific or device-specific information is used at the training stage in this work, it can be considered as a more generic approach for dealing with scenarios in which varieties of ECG signals are collected from different patients using different types of sensor devices.

## I. Introduction

The Electrocardiogram (ECG) signal is a non-invasive screening tool that has been widely used for various cardiac abnormality detections. A careful inspection of ECG signal is essential for detect underlying heart conditions particularly in long-term recordings (usually over a period of 24 hours). The recent advances in body sensor devices provide opportunities for user directed auto-diagnosis, self-monitoring and self-management services over the internet. There has been a significant increase in the number and variety of wearable ECG monitoring devices [1][2] in the last decade, which leads to the generation of massive volumes of inter-patient/device ECG signals and cloud-based services[3] for handling them. Such a trend imposes the requirements for true generic and data-driven ECG signal analysis methods and models (patient and device independent) to be developed.

The analysis of ECG signals has been extensively researched over many decades [4][5][6][7][8]. Many attempts have been made for classifying abnormal ECG beats using various methods. Even though the existing efforts do contribute as firm foundations to the domain, the problem of inter-patient ECG signal classifications has not been addressed adequately. This is due to the largely varied morphological characteristics from patient to patient, different ECG hardware implementations, and the changing measuring conditions. Traditionally, a typical machine learning based approach for ECG signal classification consists of three main steps, namely, data pre-processing, feature extraction and classification. Following such a process, various noises and artefacts (i.e. baseline wondering, muscle contraction and powerline interferences) are eliminated first [9][10]. Then a set of hand-crafted features are extracted from the pre-processed waveforms and are fed into the next classification stage. However, the time-varying dynamics and the morphological characteristics of ECG signals from different patients under different temporal and physical conditions make it difficult to extract useful features manually. For instance, even for an ECG waveform from a healthy subject, the shapes of QRS complex, P waves, and R–R intervals will not be the same from one beat to the other under different circumstances [11]. In fact, most researchers have chosen to use patient-specific data [7][9][15] at the training stage to ensure that classifiers are aware of the variances that exist in testing data, but not in the training data. This is reasonably understandable as it is almost impossible to hand-craft all features or to be confident enough that a set of pre-defined features can cover the full spectrum of any ECG signal.

In this paper, we present our work on using the densely connected convolutional neural network (DenseNet) [14] and gated recurrent unit network (GRU) [15] with attention [16] mechanism for inter-patient ECG classification problem. This work aims to evaluate and verify a hypothesis: with a deep learning model-based approach, it is possible to achieve better ECG classification performance without using either feature extraction methods or patient/device-specific information. The main contributions of this work are:

1. We developed a generic and feature-free deep learning model that outperforms the current state-of-the-art methods [7] [17] for inter-patient ECG classification.
2. We show that without changing model architecture or parameters, the performance of this method remains

Li Guo is with the School of Physical Sciences & Computing, the University of Central Lancashire, Preston, United Kingdom (e-mail:lguo@uclan.ac.uk).

Gavin Sim is with the School of Physical Sciences & Computing, the University of Central Lancashire, Preston, United Kingdom (e-mail: GRSim@uclan.ac.uk).

Bogdan Matuszewski is with the School of Engineering, the University of Central Lancashire, Preston, United Kingdom (e-mail: BMatuszewski1@uclan.ac.uk).

stable on different datasets with different properties, thus having better scalability and applicability.
3. We also compared the proposed architecture with other architectures to show its effectiveness.

The remainder of this paper is organised as follows. In section II, related work on ECG classification is surveyed and discussed, followed by a demonstration of our methodology in section III. Experiments, evaluation and results comparison are presented in section IV as well as some discussions. Conclusion and future work are given in section V.

## II. RELATED WORK

### A. Feature Engineering Based Approaches

Before deep learning became applicable, feature extraction based methods have dominated the ECG signal recognition domain over several decades. Those methods include wave shape functions [17][18][19][20], wavelet-based feature extractions [21][22][23], frequency-based feature extraction [24] and statistical feature [25]. Methods used for classifying these extracted features include support vector machines [17][26], decision trees [26], artificial neural networks [27], linear discriminants[17][18][20], self-organising maps with learning vector quantisation [28] and active learning framework [12]. However, the success of such methods often relies on the outputs from the feature extraction stage. Many of these frameworks treat ECG signals as a sequence of stochastic patterns. Therefore, complex feature extraction process and high sampling rates are required. Even though, due to the large intra-class variation, the robustness of many existing ECG classification techniques remains limited. A major limitation of current approaches is that they are highly dependent on the pre-extracted features from the training dataset, and perform inadequately while dealing with unseen ECG records. In addition, extracting complex features in the frequency domains when combined with dimensionality reduction algorithms significantly upsurge the computational complexity of the overall process. Moreover, many classifiers have not performed well in case of inter-patient variations of the ECG signals, thereby demonstrating a common shortcoming of having an inconsistent performance while classifying a new patients ECG signal.

### B. Deep Learning Based Approaches

The development of deep learning methods for feature learning[29] yields to automatic learning of good features from the raw input data. Typical deep learning architectures include deep belief networks (DBN)[30], stacked auto-encoder (SAE) [31] and convolutional neural networks (CNN)[32][33]. The success of applying deep learning in other domains, such as image recognition and language processing, has drawn the attention of ECG classification community. In the last few years, researchers have focused their efforts on using deep learning models for ECG signal analysis.

In [34], researchers have developed algorithms based on Restricted Boltzmann Machine (RBM) for two-lead heartbeat classification where the RSM model helps mine the large set of unlabelled ECG beats in the heart healthcare monitoring applications. Similar work has been carried by [35] for automatic ECG feature extraction using Deep Belief Network. For the automatic ECG classification, researchers [26] have proposed using a combination of SVM and DBN, in which DBN is used for feature learning and SVM is then applied for the classification tasks using the learned features. With the same principle, Rahlal et al.[7] has used auto-encoders for feature learning. In this work, ECG data from several cardio databases have been utilised for effective ECG feature learning. The learned features are then applied for classification on unseen data.

Along with the development of deep learning models, the more recent trend has been focused on using convolutional neural network models for direct ECG signal classification. In [36], a CNN based classification system was introduced for automatically learning feature representations from ECG data, hence eliminating the hand-crafted feature extraction stage. In another work [37], segmented ECGs are processed by an eleven-layer convolutional neural network that achieved maximum accuracy of 93.18%. Kiranyaz et al. [13] studied the patient-specific ECG monitoring system using three-layer CNN with only R-peak wave. They also result in good accuracy in the detection of supraventricular ectopic beats and ventricular ectopic beats. A deep residual neuron network [38] has been developed by [39], where a 34 layer deep CNN is applied directly, without adopting any complex pre-processing and feature engineering steps, for classifying arrhythmia. The researchers have shown that this approach achieved cardiologist-level performance using a dataset that's 500 times larger than others of its kinds.

Despite the progress the ECG community has achieved, it is evident from the literature [27][13][6][17] that researchers have used a small number of selected patient-specific data that is usually in length between 2.5 minutes and 5 minutes for model training. Although this data is then excluded from the testing data for performance evaluation, the inclusion of them in the training stage largely reduces the variants that may be encountered in some real-world applications. In other words, if we are expecting similar performance as reported in the literature when applying those methods to an application, patient-specific data needs to be collected for the "personalised models" training. While such a constraint may hold true in some of the medical setups (e.g. where 24 hours Holters are used and experts can partially label recordings), in a free-living environment where each patient only produces very short ECG waveforms (e.g. users self-monitor themselves using body sensors for less than a few minutes [40]), this does not seem to be feasible. In addition, it is particularly difficult or even impossible to ask experts to manually label small chunks of data from all stored records when the volume of data becomes overwhelming. The research presented in this paper aims to present an effective approach to mitigating this problem.

## III. METHODOLOGY

Giving the current landscape of literature, the hypophysis we hold for this work is that with careful study and a right architecture, a single deep learning model is capable of learning generic ECG waveform features, hence performing better on ECG classification tasks**.** The development of such a model should require a minimal level of prior knowledge in the traditional feature engineering/signal processing domain thus providing a neater/simpler solution. For instance, we believe there is no need to carry R-peak detection or QRS complex identification (as many of the previous work have adopted as an essential step) since they are, in fact, features from the ECG signals.

## A. The Problem

Generally, the abnormalities of the ECG signals mainly come from two aspects, namely the ill-shaped ECG beat morphology and the temporal variance between ECG beats as illustrated in Figure 1, where Fig 1.a shows regular sinus beats, Fig 1.b shows a premature beat (temporal variance between R peaks) and Fig 1.c shows ill-shaped QRS complexes (morphological variance). If a model is capable of handling both of the above cases, it needs to 1> contain a set of good enough features for reconstructing as many ECG waveforms (standard and ill-shaped) as possible; 2> have the ability for analysing sequential data. In the realm of deep learning, there are two widely used networks, namely, convolutional neural network[32] and recurrent neural network [41] with the former focusing on feature extraction and the latter on sequence analysis.

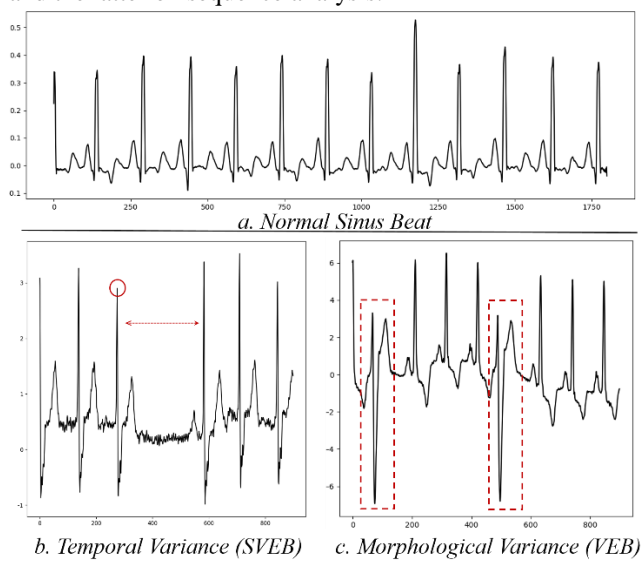

*Figure 1: Examples of ECG Beats*

There has been some research either on using stacked/deep RNN [25][29] or on using deep CNN [8][42] alone for ECG classification. It is difficult to evaluate whether these works have fully utilised the power of these models. For example, using stacked RNNs/LSTMs alone, it is based on the assumption that these stacked LSTM layers can extract good features. However, it is challenging for an LSTM layer to learn local features of input unless the model goes really deep and extensive, which ends up with almost infinite training time. In fact, the work presented in [42] has deployed wavelet sequence (WS) layers on top of the LSTM layers in their network architecture. These WS layers essentially are feature extraction layers with specific focuses on wavelet transformation. On the other hand, using CNN alone is effective in learning good local features (as well as the combinations of them). However, the time variance information is lost, and the performance of the model heavily replies on whether the model can reconstruct "close-enough" signals using features learnt. The work was done by Rajpurkar et al.[39] is precisely one of the kinds. With an extensive dataset and a deep architecture, a network surely has a better chance to learn good features and use them for classification without considering time variances. A possible bottleneck of using such approaches is that there may be a need to prepare a super-sized dataset to only use features for compensating the lost temporal information in the original inputs. Thus, it is natural to think of using a combination of both CNN and RNN for solving the ECG classification problem.

## B. The Model

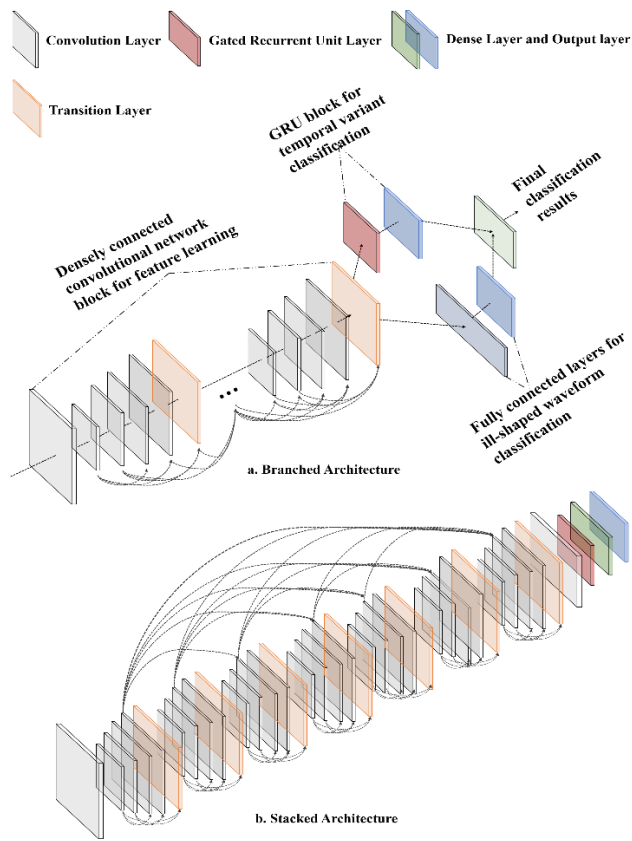

*Figure 2: Model Architecture: a) Branched b) Stacked*

We have implemented a deep learning model as illustrated in Figure 2.a. The model, in brief, has a convolutional block for feature extraction, a recurrent network block for temporal variance analysis and Softmax layers for classification. Unlike a standard stacked architecture, as shown in Figure 2.b, our model has a branched-out structure after the convolutional block. One branch is built with recurrent layers followed by a Softmax layer, and another is directly wired to a Softmax layer. Outputs from both Softmax layers are then merged with equal weights as the final output of the model. Although there have been some other efforts on using CNN and recurrent network together[43], it is conjectured that the branched architecture will work better.

For a stacked architecture, during the training process, the recurrent layers receive the full back-propagated errors. These errors are digested by them first before getting further propagated to the dense blocks. After many steps of processing, the error gets smaller and smaller before they come out from the recurrent units. The recurrent layers only have local views and try to minimise those errors without considering how the minimisation will affect the lower dense block layers. This leads to a model with better optimised top layers but less optimised lower layers. In contrast, with our branched architecture, the errors are first split equally between the recurrent block and the convolutional block. Therefore, the lower convolutional block receives higher gradients values than that of a stacked architecture, and are trained more sufficiently. The trade-off, however, is the recurrent branch receive fewer gradient values comparing to the stacked one. Since for ECG classification, we equally care about temporal variance and morphological information of a signal, the branched architecture is favoured over another one. The primary model components and the details of each are explained below.

1) *Densely connected convolutional network block*

This component is used for extracting different levels of features from the input ECG signals. DenseNet is a special kind of convolutional neural network with direct connections from one layer to all its subsequent layers. In other words, a layer in DenseNet receives outputs from all its previous layers and "*forwards*" them to the next layer together with its own output. This property is beneficial for ECG signal classifications. ECG diagnosis normally involves uses of several key clinical features, e.g. R-R interval, P wave duration, QRS duration, S-T-U duration, by cardiologists.

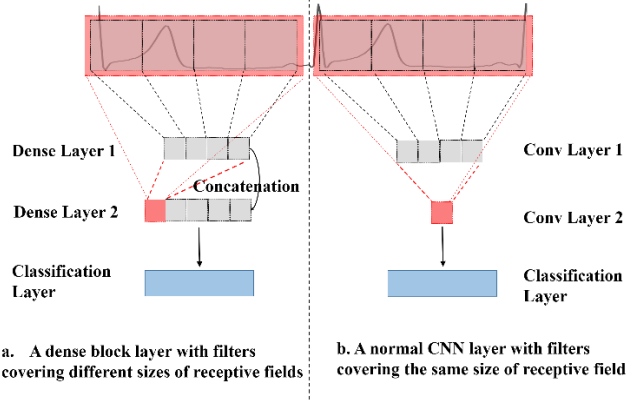

*Figure 3: DenseNet Layer vs Normal CNN Layer*

With such an architecture, low and high-level features are combined (concatenated) together at each convolutional layer (as shown in Figure 3.a). In contrast, with a "*vanilla*" version of CNN, low-level features can only contribute to the final classification through several layers of transformation (as sown in Figure 3.b). Since we do not know what feature and at what level, has a positive impact on the final classification result, it is sensible to treat them in a *flat* manner, instead of using them hierarchically. For our work, we have used dense blocks each with 4 convolutional layers inside, followed by a transition layer.

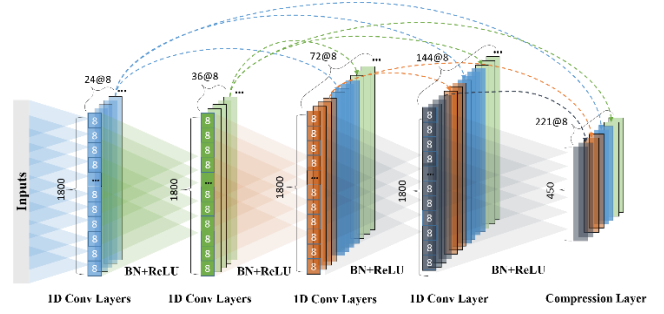

*Figure 4: Internal Look of a Dense Block. All the values in this diagram are for demonstration purpose only. In the 24@8 annotation, 24 is the filter number and 8 is the filter size used for that conv layer. The example growth rate used for this dense block is 12, which means the successive conv layer always has 12 more filters in addition to the concatenated filters from its last two layers. Taking the 3$^{rd}$ conv layer as an example, the concatenated filters are 60 from the last two layers. As the growth rate is 12, it has 70 filters in total. The growth rate does not apply to the last transition layer. For a compression rate of 0.8, it has 221 (276*0.8) filters.*

Figure 4 shows the internal look of a dense block. With such a structure, filters learnt from the previous layers are pushed towards to the end of the chain along with the new filters from the later layers. Outputs from each layer are batch normed[44], first for removing covariate shifts before they go into rectified linear units (ReLU) for activations. The transition layer (the last layer coloured in yellow in the diagram) plays a compression role to ensure the size of the network does not explode. It takes activations from all the filters in a dense block and compresses them by using a smaller number of filters and pooling operations.

2) *Gated recurrent unit block with attention*

This component receives activations from the last dense block and analyses them sequentially for learning temporal features of the inputs. The key elements in this component are the gated recurrent network layers. GRU can be considered as a variation of long-short-term-memory(LSTM) [45] because both are designed similarly and, in some cases, produce equally excellent results[15]. For this work, we choose GRU purely based on the consideration of training time. As the GRU does not introduce the forget gate, it reduces the computation time accordingly.

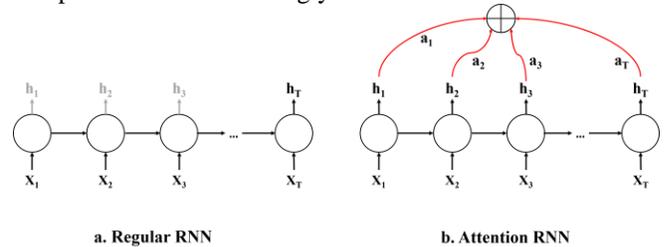

*Figure 5: Gated Recurrent Unit with Attention*

Inspired by the work from image recognition and the language translation domain, we also implemented the attention mechanism[16] for the GRU layer. The attention mechanism, in short, is to make the GRU layer focus on and learn (pay more attention) specific parts/features of the inputs and produce results accordingly. Taking the two abnormal ECG waves from Figure 1 as examples, when a standard GRU layer sees them, every data point in the signal is analysed one by one sequentially, and the final state of the GRU is used for producing the final results. More specifically, the network only uses the final state of the GRU for its output as shown in Figure 5.a. An attention enabled GRU unit as shown in Figure 5.b, instead, takes the sum of the weighted combination of all the GRU internal states and use it for final classification. Therefore, the most significant parts of the input (in our case, the ill-shaped waves) receives more updates at the network back-propagation stage.

## IV. EXPERIMENTS AND EVALUATION RESULTS

### A. Dataset

To work with real device independent and inter-patient data, we chose to use ECG records from two open-source PhysioBank databases, namely the MIT-BIH Arrhythmia database and the MIT-BIH Supraventricular Arrhythmia database [46].

#### 1) The MIT-BIH Arrhythmia Database (mitdb)

This database consists of a 48-half-hour long ECG recording from 47 patients. Each record is sampled at 360Hz and was interpreted, validated and annotated by at least two cardiologists. This dataset contains 23 recordings that were randomly selected from a set of 4000 ambulatory 24-hours ECGs and were collected from a mixed cohort of inpatients and outpatients at the medical centre. The other 25 recordings were selected from the same 4000 set which includes less common but clinically symbolic arrhythmias. Signals from mitdb have two channels, and we used data from the MLII[46] channel for generating our datasets.

#### 2) The MIT-BIH Supraventricular Arrhythmia database (svdb)

This database includes 78 half-hour ECG recordings chosen to supplement the examples of supraventricular arrhythmias in the MIT-BIH Arrhythmia Database. Records in this database are all sampled at 128Hz and have two channels as well. ECG1 [46] channel is used in this work.

#### 3) Classes Mapping to AAMI Standard

To compare the classification results with the state-of-the-art works, the preparation of dataset closely follows the recommendations of the Association for the Advancement of Medical Instrumentation (AAMI) [47] for class labelling and results in the presentation. More specifically, the AAMI standard defines five classes of interest: normal (N), ventricular (VEB), supraventricular (SVEB), fusion of normal and ventricular (F) and unknown beats (Q). Regardless of the class definition, this standard essentially recommends for performance evaluation of inter-patient scenario (i.e., training and test ECG beats are extracted from different patients). The ECG records from the MIT-BIH database for example, however, have 15 different beat types and therefore need to be mapped to these five classes accordingly. The beat type mappings between the AAMI recommended class types and MIT-BIH beat type are shown in Table 1.

| AAMI Heartbeats Classes | N | SVEB | VEB | F | Q |
|---|---|---|---|---|---|
| Description | Any heartbeat not in S,V, F or Q class | Supraventricular ectopic beat | Ventricular ectopic beat | Fusion beat | Unknown beat |
| MIT-BIH Heartbeat types (annotation symbols) | Normal (N) | Atrial premature beat (A) | Premature ventricular contraction (V) | Fusion of ventricular & normal beat (F) | Paced beat (/) |
| | Left bundle branch block beat (L) | Aberrated atrial premature beat (a) | Ventricular escape beat (E) | | Fusion of paced and normal beat (f) |
| | Right bundle branch block beat (R) | Nodal (Junctional) premature beat (J) | | | Unclassified beat (Q) |
| | Atrial escape beat (e) | Supraventricular premature beat (S) | | | |
| | Nodal (Junctional) beat (j) | | | | |

*Table 1: Mapping the MIT-BIH Arrhythmia Database Heartbeat Types to the AAMI Heartbeat Classes*[17]

### B. Data Pre-processing and Segmentation

Baseline-wondering of all the ECG records is first removed using 200ms and 600ms median filter, followed by applying a uniform moving average with window size 7 for removing high-frequency powerline and muscle noises. Finally, to speed up the training process, all records are resampled from their original sample rate (360Hz for mitdb, 128Hz for svdb) to 180Hz.

To ensure the model to be trained with enough information of both temporal variance and morphological structures, we choose to use 10 seconds length data segmentation that contains at least 6 heartbeats (40 heartbeats per minute for the worst case scenario), as the input for our model. Training data segments are generated by taking 10 seconds data points from ECG records using their beat annotation indexes (shifted one by one from the beginning). The corresponding target data is encoded using the one-hot-encoding vector. The class label of target data is determined by the most occurred annotation in the corresponding training data segment. N annotations are excluded if annotations of other classes exist.

In order to compare the performance of our work with the others', data segments from a 22 patients subset (referred as $DS_1$ later) is chosen as the first training set: $DS_1$ = { 101, 106, 108, 109, 112, 114, 115, 116, 118, 119, 122, 124, 201, 203, 205, 207, 208, 209, 215, 220, 223, 230}, while segments from another subset (referred as $DS_{T1}$) of 22 patients are chosen as the testing set: $DS_{T1}$ ={ 100, 103, 105, 111, 113, 117, 121, 123, 200, 202, 210, 212, 213, 214, 219, 221, 222, 228, 231, 232, 233, 234}. The remaining 4 patient recordings from the *mitdb* are not considered as they are on pacemakers and are consisted of only paced, (unknown type) heartbeats. We also generated another training set $DS_2$ and testing set $DS_{T2}$ that includes all the records from the *svdb*. As a result, we generated 51,912 samples from $DS_1$, 92,724 samples from $DS_2$ for training, and 50,900 samples from $DS_{T1}$, 94,130 samples from $DS_{T2}$ for testing. Table 2 shows the breakdown of the 5 classes of beat subtypes in the generated datasets. Training and testing datasets are standardised with a mean of 0 and a standard deviation of 1 separately.

|       | N       | SVEB   | VEB    | F     | Q   | Total   |
|-------|---------|--------|--------|-------|-----|---------|
| DS$_1$    | 30,966  | 4,854  | 15,648 | 409   | 35  | 51,912  |
| DS$_{T1}$  | 29,980  | 4,225  | 15,373 | 1,251 | 71  | 50,900  |
| DS$_2$    | 39,669  | 30,064 | 22,672 | 137   | 182 | 92,724  |
| DS$_{T2}$  | 48,690  | 22,997 | 22,329 | 17    | 97  | 94,130  |
| Total | 149,305 | 62,140 | 76,022 | 1,814 | 385 | 289,666 |

*Table 2: Breakdown of Beat Subtypes in the Generated Dataset*

### C. Evaluation Metrics, Model Parameters and Training

The measurement metrics used for evaluating classification performance are as follows:

Accuracy: Acc = (TP+TN)/(T P+FP+TN+FN),

Sensitivity: Sen = TP/(TP+FN),

Specificity: Spe=TN/(FP+TN),

Positive Predictivity: Ppr=TP /(FP+TP).

The above four metrics are computed by the quantity of true positive (TP), false positive (FP), true negative (TN), and false negative (FN). $F_1$ Score=2(Sen*Ppr) /(Sen+Ppr) is used as the combined metric for performance comparison.

There are a few key model parameters that need to be set before model training. Some of them are chosen using specific clinical knowledge for ECG analysis, and some of them are set via parameter searching. Each of the four strategies is explained below:

1. *Filter/kernel size for convolutional layers:* Filter size for the initial convolutional layer is set to 8 and remains unchanged in all convolutional layers. This is because, for our 180Hz sampled data segment, 8 data points represent 44ms (5.6ms each data point) in time that is sufficiently small for composing most of the ECG features (e.g. QRS complexes normally have the shortest duration, around 60ms).

2. *Number of convolutional layers in a dense block:* Due to the concatenation operations required in the dense blocks, the first dimension (signal length) of adjacent layers have to be the same. Therefore, it is not possible to perform pooling operations inside a dense block, which means the stride size for each convolutional layer has to be exactly one. As a consequence, the depth of a dense block actually is determined by the filter size of the final convolutional layer inside it. Since a regular P-R interval from an ECG QRS complex has the largest value (around 120ms-200ms) amongst all the important ECG waves and intervals, for a single filter that can cover this, we need to make the size of it to 40 data samples. This, as a result, leads to a dense block with 5 convolutional layers (filter size 8 for the first layer and 40 for the last layer).

3. *Input length for the GRU layers:* Since GRU layers use outputs from the dense block components and use them for time variance analysis. It is rational to decide the input length for GRU layer first as this decision will affect how many dense blocks we will need on the top. For a 10 seconds ECG segment, we know there are around 8-30 QRS complexes. For the extreme case (28 QRS complexes), we will need 28 input features to cover the full length. The input length for the GRU layer is therefore set to 28. This also indicates that the output from the lower dense block has to be 28.

4. *Pool size and pool stride for transition layers:* Because we are not able to perform pooling operations inside a dense block, it has to be done at the transition layer. To reduce the sequence length from 1800 (180*10) to 30, we will need 6 pooling operations for the stride value of 2, so are 6 transition layers and 6 dense blocks. Alternatively, we could have 4 pooling operations with the stride value of 4, which leads to 4 dense blocks and 4 transition layers. In order to determine the best stride value for all transition layers, we carried a simple parameter search experiment in which we used a small dataset (10% class-balanced data from DS$_1$) for training models with all possible stride value ranging from 2 to 8 and used the $F_1$ score for SVEB classification as the performance metric. The result shows that when we set the stride value to 2 in all transition layers, the model achieved the best performance at 51.23, whereas 50.13 when the stride value is 4. We choose to use the latter one in our final model as giving the close performance, a larger stride value helps reduce the total model parameters.

As we applied batch normalisation after each convolution layer, a small dropout value 0.25 is adopted for model regularisation. We train our model for 500 epochs with a batch size of 50 using an early stop monitor on validation loss with 50-epochs patience. The optimiser we use is Adam[48] with an initial learning rate set to 5e-4. The training is carried on NVIDIA GTX 1080i GPUs and the average training time for each epoch of the full DS$_1$ training set is around 230 seconds. Table 3 summarises the values of key model parameters used in this study.

| Parameter | Values |
|---|---|
| **Filter size of conv layers:** | 8 |
| **No. of filters & growth rate for conv layers:** | 48&24 |
| **No. of conv layers in each dense block:** | 4 |
| **No. of dense blocks:** | 4 |
| **Transition layer pool size:** | 4 |
| **Transition layer stride size:** | 4 |
| **Transition layer compression rate:** | 0.8 |
| **Input sequence length of GRU layer:** | 28 |
| **Units of GRU layer:** | 64 |
| **Dropout value:** | 0.25 |
| **Mini-batch size:** | 50 |
| **Initial learning rate:** | 5e-4 |

*Table 3: Values of Key Model Parameters*

### D. Model Performance Evaluation

To evaluate the performance of the proposed work, we compare the results of our work with existing methods that also comply with the AAMI standard. According to AAMI recommendations, the SVEB and VEB detections are considered separately. The performance evaluation results for all our experiments are given in Table 4. The works [5][6][22] that have used a good number of annotated beats from the testing records for training purposes are not included in this study. Although those beats are not used in the testing stage, they do help the model learn patient or even record specific patterns thus leading to a better performance when given the rests of the "*already seen*" record.

| Methods | SVEB | | | | | VEB | | | | |
|---|---|---|---|---|---|---|---|---|---|---|
| | **$T_1$**: Classification results using all training data in $DS_1$ and all testing data $DS_{T1}$ (mitdb) | | | | | | | | | |
| | Acc (%) | Sen (%) | Spe (%) | Ppv (%) | $F_1$ | Acc (%) | Sen (%) | Spe (%) | Ppv (%) | $F_1$ |
| Chazel et al. [17] | 94.6 | 75.9 | N/A | 38.5 | 51.08 | 97.4 | 77.7 | N/A | 81.9 | 79.74 |
| Matthew et al. [49] | 93.4 | 75.12 | N/A | 32.84 | 45.7 | 93.51 | 76 | N/A | 49.97 | 65.13 |
| Matthew et al.[4] | 93.78 | 88.39 | N/A | 33.63 | 48.72 | 96.63 | 77.74 | N/A | 69.2 | 80.53 |
| Rahlal et al. [7] | 94.9 | 37.8 | 97.5 | 40.5 | 39.1 | 97.8 | 90.1 | 98.6 | 87.1 | 88.57 |
| Jiang el al [27] | 96.6 | 50.6 | 98.8 | 67.9 | 57.99 | 98.1 | 86.6 | 99.3 | 93.3 | 89.82 |
| Ince el al.[21] | 96.1 | 62.1 | 98.5 | 56.7 | 59.28 | 97.6 | 83.4 | 98.1 | 87.4 | 85.21 |
| Raj et al.[39]* | 89.79 | 38.79 | 94.41 | 38.58 | 38.6 | 87.12 | 88.57 | 88.57 | 73.93 | 80.59 |
| Stacked* | 93.82 | 72.99 | 95.71 | 60.61 | **66.23** | 90.08 | 89.57 | 90.30 | 79.98 | 84.50 |
| This work | 93.61 | 62.70 | 96.40 | **61.21** | 61.94 | 93.71 | **91.25** | 94.77 | **88.30** | **89.75** |
| | **$T_2$**: Classification results using random training data in $DS_1$&$DS_2$ and all testing data in $DS_{T1}$ (mitdb) | | | | | | | | | |
| This work | 92.54 | 69.82 | 94.60 | 53.92 | 60.85 | 93.54 | 91.23 | 94.54 | 87.85 | 89.51 |
| | **$T_{3.a}$**: Classification results using the full mitdb ($DS_1$+$DS_{T1}$) for training and testing data in $DS_{T2}$ (svdb) | | | | | | | | | |
| Rahlal et al. [7] | 90.61 | 8.80 | 96.32 | 14.31 | 10.90 | 66.27 | 65.19 | 66.32 | 6.31 | 11.51 |
| This work | 68.73 | 7.90 | 97.91 | 64.50 | 14.08 | 81.92 | 86.81 | 80.33 | 58.82 | 70.13 |
| | **$T_{3.b}$**: Classification results using pre-trained model from $T_2$ and all testing data in $DS_{T2}$ (svdb) | | | | | | | | | |
| This work | 88.54 | 71.29 | 94.12 | 79.66 | 75.24 | 92.17 | 79.96 | 95.96 | 86.03 | 82.89 |
| | **$T_4$**: Classification results using pre-trained model from $T_2$ and all testing data in $DS_{T1}$ & $DS_{T2}$ (mitdb+svdb) | | | | | | | | | |
| This work | 88.99 | 69.35 | 94.38 | 77.20 | 73.07 | 91.71 | 81.80 | 95.93 | 89.52 | 85.49 |

*Table 4: Classification results in terms of SVEB and VEB using the testing records of MIT-BIH. **1)** Work that is marked with * is implemented by us for comparison purposes as we are interested in them but cannot find literature that uses the same testing dataset. For instance, the original work in Raj et al. [46]* has used the PhysioNet/CiC Challenge 2017 dataset for testing. To evaluate the effectiveness of the model, we had to implement the model as described in the original paper and applied it to our testing dataset with a few parameter tunings. However, this should not de-value the author's original work. Stacked* is the model shown in Figure 2.b and is fine-tuned using similar approaches applied to the branched architecture. **2)** The two work that is greyed out has used 300 beats for each record from the testing set.*

*1) Performance Evaluation using mitdb Data*

To evaluate how training data size affects the model performance, the model is first trained with datasets of different sizes from $DS_1$ and then evaluated using the full testing records $DS_{T1}$. Training data is randomly drawn from $DS_1$ with balanced loads for all classes with a gradual size increase in the dataset. Performances of the model for different training data sizes are given in Figure 6. It can be seen that the overall performance of the model does increase along with the training data size except for the F class which is largely under-represented in $DS_1$. The model achieved its best performance for SVEB and F classes ($F_1$=**65.49** for SVEB class and $F_1$=**60.73** for F class) with training datasets of size 40946, whereas for Normal class and VEB class, the best performance is achieved with the full $DS_1$ data ($F_1$=**90.54** for SVEB class and $F_1$=**89.75** for F class). There is a noticeable performance drop on the chart for SVEB and F classes with full training data $DS_{T1}$.

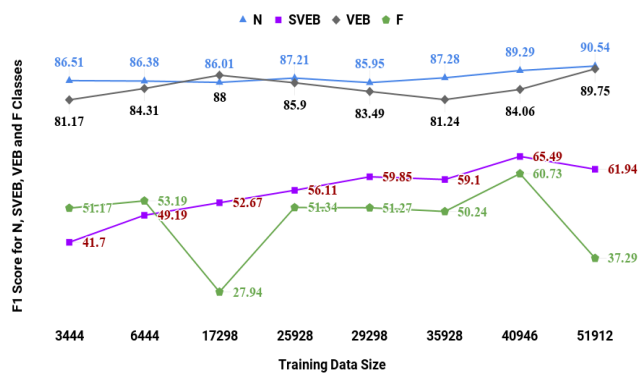

*Figure 6: Classification Performances with Different Sizes of Training Data from $DS_1$. The above performance figure for each data size is the aggregated mean of multiple testing results using non-replacement data sampling. As the data size increases, when the required sample number of a class exceeds its maximum in $DS_1$, the non-replacement sampling only applies to the rests of classes. This process is repeated until the data are fully sampled from the $DS_1$ in a single go.*

As shown in Table 4, for experiment $T_1$, this proposed work (branched architecture) has achieved considerable performance improvement on SVEB detection. The $F_1$ score is improved by almost 21% from 51.08 (the current state-of-the-art) to **66.23/61.94** (stacked/branched). It also performs better on the VEB detection with ($F_1$=**89.75**) than the current state-of-the-art performance ($F_1$=**88.57**). Even for the work [21][27] that have used beats from the testing data in their training, our model still outperforms on SVEB detection and with a very close performance on VEB detection ($F_1$=89.75 vs $F_1$=89.82).

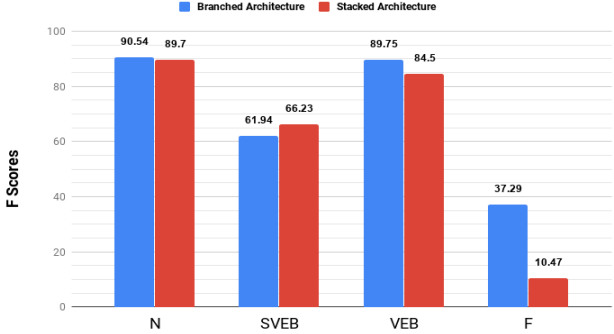

*Figure 7: Classification Performances for 4 classes (Stacked vs Branched).*

It can be noticed that from the results, the stacked architecture achieved the highest score on SVEB detection with $F_1$=66.23. However, if we look at the overall classification performances, clearly the branched architecture performs better. From Figure 7, it can be seen that the stacked network is only more effective on SVEB detection, whereas it performs worse on N, VEB and F classes. This is consistent with the discussion in section III.2.

*2) Model Performance and Scalability Evaluation Using mitdb and svdb Data*

To verify the scalability of the model, additional training and testing of the model was conducted using both mitdb and svdb data.

Experiment $T_2$ aims to test that with a relatively larger, more balanced but heterogeneous dataset, whether there would be improvements in the model performance or not. Firstly, 5 new training datasets were constructed, each with the 51763 data via non-replacement sampling from $DS_1$ (mitdb) and $DS_2$ (svdb). They represent better balanced datasets (N: 17,000, SVEB: 17,000, VEB: 17,000, F: 546, Q: 217)[1]. Following the first stage, the model with the best performance from $T_1$ was re-trained using the new datasets and was tested using the $DS_{T1}$ dataset. The test result shows that the model averagely scored $F_1$= 92.67 for N class, $F_1$= 60.85 for SVEB class, $F_1$= 89.51 for VEB class and $F_1$= 62.65 for F class. The results are similar to what the model achieved using the $DS_1$ only dataset for training, with clear improvement on N and F class detection. Also, an interesting observation of this experiment is that given the model performances for SVEB and VEB classification remain consistent, this reveals that the mitdb and svdb data do share some common features.

---
[1] Unless otherwise stated, all the data in the new training datasets are resampled to 180Hz and re-standardised.

In $T_{3.a}$, we re-trained the model from $T_1$ using the full mitdb data ($DS_1$+$DS_{T1}$) and applied the final model to the $DS_{T2}$ (svdb) testing data. In this experiment, the model has not seen any svdb data in training and can only use features learnt from the mitdb for classifications. Even for such an extreme case, it has successfully (comparing to the work presented in [7]) scored $F_1$=70.13 for VEB detection. This can further prove that the model has learnt generic morphological features. Despite the fact that the SVEB detection result is also better than another work [7] ($F_1$=14.08 vs $F_1$=10.90), the low value indicates that the temporal variant information is very data dependent and cannot be easily generalised from a small dataset with varied properties. In $T_{3.b}$, we tested the pre-trained model from $T_2$ using $DS_{T2}$ testing data. We use this experiment to simulate a practical scenario where a pre-trained model is used for classifying new data. The model performance was boosted to $F_1$=75.24 for SVEB detection and $F_1$=82.89 in this case.

In the last experiment $T_4$, the scalability of the model was investigated further via testing the pre-trained model from $T_2$ on classifying the full testing data from $DS_{T1}$ and $DS_{T2}$. The results (N: 92.70, SVEB: 73.07, VEB: 85.49, F: 59.85) are, in fact, very close to the weighted (based on the number of class samples in the testing data from each database) averages of $T_2$ and $T_{3.b}$'s results. This suggests that the model's performance and stability are not affected by the heterogeneities of different data sources, but are only dependent on whether such heterogeneities are represented sufficiently in the training data.

*E. Discussion*

Although from the experiments, the proposed architecture has been approved to deliver consistent performance, it is still interesting and necessary for us to understand what the model has learnt. To gain insight into the model, it is necessary to visualise the activations of the last Conv layer before the GRU layer, and the activations of the GRU layer after the attention mechanism is applied.

Figure 9 shows the activations of a group of correctly classified and misclassified input signals. The high spikes in the activation map show what the model focus on when it sees the signal. For those correctly classified samples, we can see that both conv layer and GRU layer have the right focus (high spikes at the right position), while the GRU layer seems to be more detailed. For example, for the a.1 signal, we have two noticeable spikes in the conv layer activation map but four in the GRU's one that covers the four ill-shaped beats in the input signal (S, V, V, V). This is similar to the a.2 case. However, for the a.3 case, the conv layer seems to be more sensible as it spikes both F beats whereas the GRU layer only flags the latter one with a more focused spike. It is also interesting to see both layers treat a.4 as an N record despite that it has a V shape noise in the middle. As for the misclassified sample, b.1, the conv layer successfully spikes it, but the GRU layer almost completely ignores it. For b.2, both layers focus at the right position but not with enough confidence (the spikes are not significant enough). b.3 is an interesting case, both layers spiked it successfully but considered it as V class. The F beat does look like a V beat as shown in a.2. However, the network manages to classify the

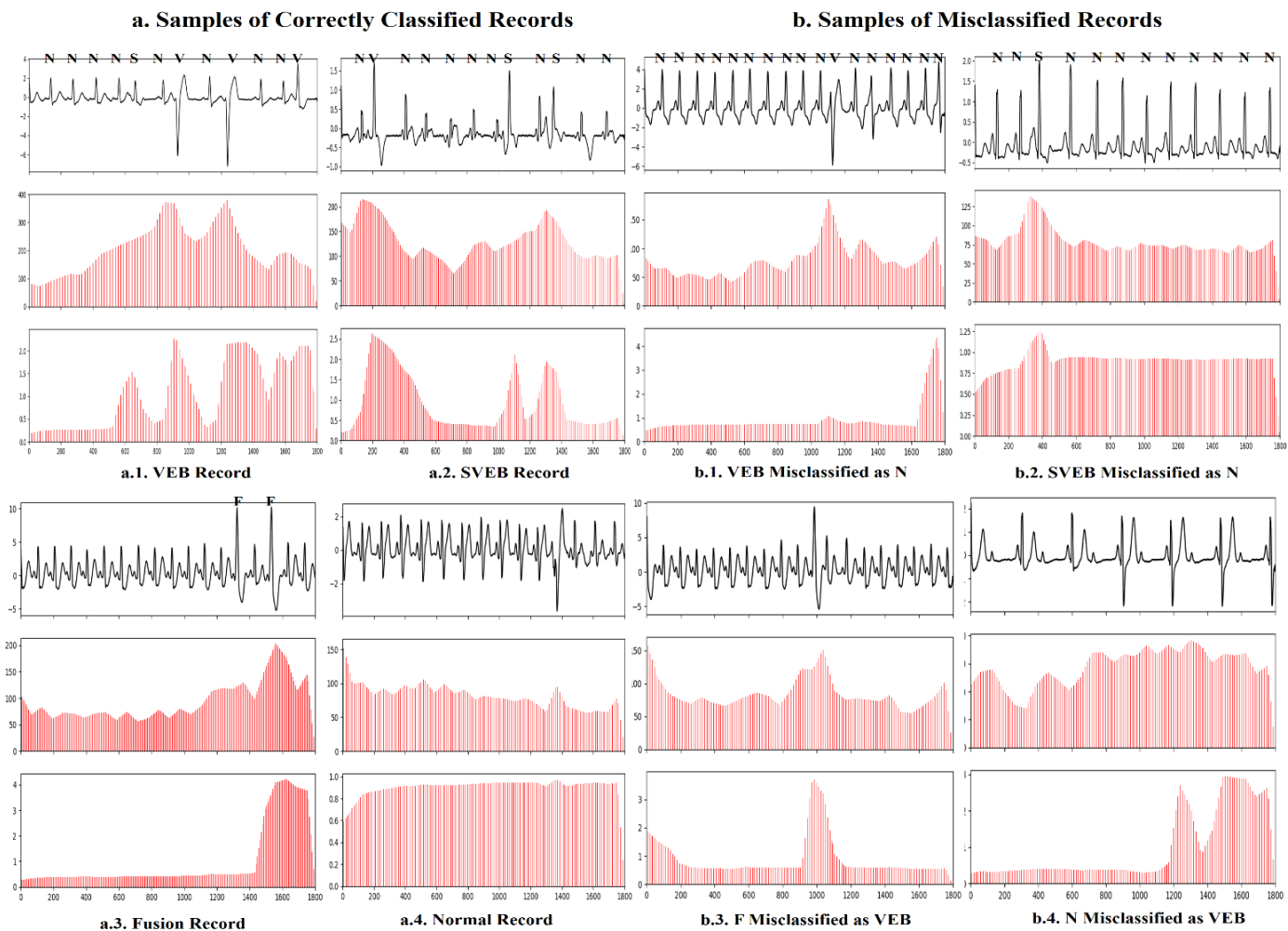

*Figure 8: Visualisation of Layer Activations. The **first row** shows the input ECG signal. The **second row** shows layer activations of the last convolution layer before the GRU layer. The **third row** shows layer activations of the GRU layer. Each bin in the diagram represent the sum of activations for all filters/units that cover a specific local area in the original signal. Putting all them together, we are able to visualise the total activations from a layer for all different local areas in the input signal. In other words, a local area that receives more activations (high spikes in the figure) is considered more important by the network for the final classification. Note: As we have applied pooling operations in the model (input size gets reduced from 1800 to 28), the above activation map is re-constructed (from 28 back to 1800) using un-pooling operations.*

a.2 correctly which contains two similar shaped F beats. This requires further investigation and understanding of the network behaviour. The misclassification for b.4 is easy to understand. The last four beats, although annotated as normal beats, have different shapes as compared to the first two normal beats. Looking into the activation map, it is believed that some of the misclassifications can be improved via further parameter tuning, e.g. filter size in particular and larger network capacity. This will be the basis for future work.

## V. CONCLUSION AND FUTURE WORK

As we are entering the cloud-enabled healthcare era, the development of automatic diagnosis/early warning services based on large volume but heterogeneous data sources has become the key to success.

In this paper, we presented our work on using a branched deep learning model for inter-patient ECG classification problem. We discussed the model design rationale and justified out choices of each component. How the important parameters of the model are chosen is also explained. The experiments conducted show that this proposed model has achieved considerable performance improvement compared to the current state-of-the-art with good scalability. The experiment results also indicate that the model performance remains consistent across two heterogeneous data sources.

To further extend this work, we will increase the depth of the network to test if the model can learn more generic features (features that are useful for any ECG dataset). We also plan to apply a similar approach to multi-label ECG classification problem. In this work, data segments are annotated using the maximum occurred annotations from the databases. However, in reality, it is often to see records that contain multiple classes (as shown in Figure 8.a.1 and a.2). A model that is able to label all different types of classes accurately would undoubtedly help with the fast developing trend of cloud-based healthcare services.

## References

[1] H. Kim, S. Kim, N. Van Helleputte, A. Artes, M. Konijnenburg, J. Huisken, C. Van Hoof, and R. F. Yazicioglu, 'A Configurable and Low-

Power Mixed Signal SoC for Portable ECG Monitoring Applications', *IEEE Trans. Biomed. Circuits Syst.*, vol. 8, no. 2, pp. 257–267, 2014.
[2] D. D. Mehta, N. T. Nazir, R. G. Trohman, and A. S. Volgman, 'Single-lead portable ECG devices: Perceptions and clinical accuracy compared to conventional cardiac monitoring', *J. Electrocardiol.*, vol. 48, no. 4, pp. 710–716, Jul. 2015.
[3] Y. Li, L. Guo, and Y. Guo, 'Enabling Health Monitoring as a Service in the Cloud', in *2014 IEEE/ACM 7th International Conference on Utility and Cloud Computing*, 2014, pp. 127–136.
[4] S. M. Mathews, C. Kambhamettu, and K. E. Barner, 'A novel application of deep learning for single-lead ECG classification', *Comput. Biol. Med.*, vol. 99, pp. 53–62, Aug. 2018.
[5] P. De Chazal, M. O'Dwyer, and R. B. Reilly, 'Automatic classification of heartbeats using ECG morphology and heartbeat interval features', *IEEE Trans. Biomed. Eng.*, vol. 51, no. 7, pp. 1196–1206, 2004.
[6] Y. H. Hu, S. Palreddy, and W. J. Tompkins, 'A patient-adaptable ECG beat classifier using a mixture of experts approach', *IEEE Trans. Biomed. Eng.*, vol. 44, no. 9, pp. 891–900, 1997.
[7] M. M. Al Rahhal, Y. Bazi, H. AlHichri, N. Alajlan, F. Melgani, and R. R. Yager, 'Deep learning approach for active classification of electrocardiogram signals', *Inf. Sci. (Ny).*, vol. 345, pp. 340–354, Jun. 2016.
[8] S. Chauhan and L. Vig, 'Anomaly detection in ECG time signals via deep long short-term memory networks', in *2015 IEEE International Conference on Data Science and Advanced Analytics (DSAA)*, 2015, pp. 1–7.
[9] B. H. Tracey and E. L. Miller, 'Nonlocal means denoising of ECG signals', *IEEE Trans. Biomed. Eng.*, vol. 59, no. 9, pp. 2383–2386, 2012.
[10] R. Sameni, M. B. Shamsollahi, C. Jutten, and G. D. Clifford, 'A nonlinear Bayesian filtering framework for ECG denoising', *IEEE Trans. Biomed. Eng.*, vol. 54, no. 12, pp. 2172–2185, 2007.
[11] R. Hoekema, G. J. H. Uijen, and A. Van Oosterom, 'Geometrical aspects of the interindividual variability of multilead ECG recordings', *IEEE Trans. Biomed. Eng.*, vol. 48, no. 5, pp. 551–559, 2001.
[12] J. Wiens and J. V Guttag, 'Active learning applied to patient-adaptive heartbeat classification', in *Advances in neural information processing systems*, 2010, pp. 2442–2450.
[13] S. Kiranyaz, T. Ince, and M. Gabbouj, 'Real-Time Patient-Specific ECG Classification by 1-D Convolutional Neural Networks', *IEEE Trans. Biomed. Eng.*, vol. 63, no. 3, pp. 664–675, 2016.
[14] G. Huang, Z. Liu, L. van der Maaten, and K. Q. Weinberger, 'Densely Connected Convolutional Networks', Aug. 2016.
[15] J. Chung, C. Gulcehre, K. Cho, and Y. Bengio, 'Empirical Evaluation of Gated Recurrent Neural Networks on Sequence Modeling', Dec. 2014.
[16] K. Gregor, I. Danihelka, A. Graves, D. J. Rezende, and D. Wierstra, 'DRAW: A Recurrent Neural Network For Image Generation', Feb. 2015.
[17] P. De Chazal, M. O'Dwyer, R. B. R. B. Reilly, P. deChazal, M. O'Dwyer, and R. B. R. B. Reilly, 'Automatic classification of heartbeats using ECG morphology and heartbeat interval features', *IEEE Trans. Biomed. Eng.*, vol. 51, no. 7, pp. 1196–1206, Jul. 2004.
[18] P. de Chazal and R. B. Reilly, 'A Patient-Adapting Heartbeat Classifier Using ECG Morphology and Heartbeat Interval Features', *IEEE Trans. Biomed. Eng.*, vol. 53, no. 12, pp. 2535–2543, Dec. 2006.
[19] P. de Chazal, 'Detection of supraventricular and ventricular ectopic beats using a single lead ECG', in *2013 35th Annual International Conference of the IEEE Engineering in Medicine and Biology Society (EMBC)*, 2013, pp. 45–48.
[20] M. Llamedo and J. P. Martínez, 'Heartbeat Classification Using Feature Selection Driven by Database Generalization Criteria', *IEEE Trans. Biomed. Eng.*, vol. 58, no. 3, pp. 616–625, Mar. 2011.
[21] T. Ince, S. Kiranyaz, and M. Gabbouj, 'A Generic and Robust System for Automated Patient-Specific Classification of ECG Signals', *IEEE Trans. Biomed. Eng.*, vol. 56, no. 5, pp. 1415–1426, May 2009.
[22] J. L. Herlocker and J. A. Konstan, 'Content-independent task-focused recommendation', *Internet Comput. IEEE*, vol. 5, no. 6, pp. 40–47, 2001.
[23] X. Jiang, L. Zhang, Q. Zhao, and S. Albayrak, 'ECG Arrhythmias Recognition System Based on Independent Component Analysis Feature Extraction', in *TENCON 2006 - 2006 IEEE Region 10 Conference*, 2006, pp. 1–4.
[24] L. Senhadji, G. Carrault, J. J. Bellanger, and G. Passariello, 'Comparing wavelet transforms for recognizing cardiac patterns', *IEEE Eng. Med. Biol. Mag.*, vol. 14, no. 2, pp. 167–173, 1995.
[25] G. de Lannoy, D. Francois, J. Delbeke, and M. Verleysen, 'Weighted Conditional Random Fields for Supervised Interpatient Heartbeat Classification', *IEEE Trans. Biomed. Eng.*, vol. 59, no. 1, pp. 241–247, Jan. 2012.
[26] J. Rodriguez, A. Goni, and A. Illarramendi, 'Real-Time Classification of ECGs on a PDA', *IEEE Trans. Inf. Technol. Biomed.*, vol. 9, no. 1, pp. 23–34, Mar. 2005.
[27] W. Jiang, S. G. Kong, Wei Jiang, and G. Seong Kong, 'Block-based neural networks for personalized ECG signal classification', *IEEE Trans. Neural Networks*, vol. 18, no. 6, pp. 1750–1761, Nov. 2007.
[28] M. Lagerholm, C. Peterson, G. Braccini, L. Edenbrandt, and L. Sornmo, 'Clustering ECG complexes using Hermite functions and self-organizing maps', *IEEE Trans. Biomed. Eng.*, vol. 47, no. 7, pp. 838–848, Jul. 2000.
[29] Y. Bengio, A. Courville, and P. Vincent, 'Representation learning: A review and new perspectives', *IEEE Trans. Pattern Anal. Mach. Intell.*, vol. 35, no. 8, pp. 1798–1828, 2013.
[30] G. E. Hinton, S. Osindero, and Y.-W. Teh, 'A fast learning algorithm for deep belief nets', *Neural Comput.*, vol. 18, no. 7, pp. 1527–1554, 2006.
[31] P. Vincent, H. Larochelle, Y. Bengio, and P.-A. Manzagol, 'Extracting and composing robust features with denoising autoencoders', in *Proceedings of the 25th international conference on Machine learning*, 2008, pp. 1096–1103.
[32] Y. LeCun and Y. Bengio, 'Convolutional networks for images, speech, and time series', *Handb. brain theory neural networks*, vol. 3361, no. 10, p. 1995, 1995.
[33] R. Donida Labati, E. Muñoz, V. Piuri, R. Sassi, and F. Scotti, 'Deep-ECG: Convolutional Neural Networks for ECG biometric recognition', *Pattern Recognit. Lett.*, Mar. 2018.
[34] Y. Yan, X. Qin, Y. Wu, N. Zhang, J. Fan, and L. Wang, 'A restricted Boltzmann machine based two-lead electrocardiography classification', in *2015 IEEE 12th International Conference on Wearable and Implantable Body Sensor Networks (BSN)*, 2015, pp. 1–9.
[35] D. Wang and Y. Shang, 'Modeling Physiological Data with Deep Belief Networks.', *Int. J. Inf. Educ. Technol.*, vol. 3, no. 5, pp. 505–511, 2013.
[36] M. Zubair, J. Kim, and C. Yoon, 'An Automated ECG Beat Classification System Using Convolutional Neural Networks', in *2016 6th International Conference on IT Convergence and Security (ICITCS)*, 2016, pp. 1–5.
[37] U. R. Acharya, H. Fujita, S. L. Oh, U. Raghavendra, J. H. Tan, M. Adam, A. Gertych, and Y. Hagiwara, 'Automated identification of shockable and non-shockable life-threatening ventricular arrhythmias using convolutional neural network', *Futur. Gener. Comput. Syst.*, vol. 79, pp. 952–959, Feb. 2018.
[38] K. He, X. Zhang, S. Ren, and J. Sun, 'Deep Residual Learning for Image Recognition', Dec. 2015.
[39] P. Rajpurkar, A. Hannun, … M. H. preprint arXiv, and undefined 2017, 'Cardiologist-level arrhythmia detection with convolutional neural networks', *arxiv.org*.
[40] D. Albert, B. R. Satchwell, and K. N. Barnett, 'Wireless, ultrasonic personal health monitoring system'. Google Patents, 2012.
[41] T. Mikolov, M. Karafiát, L. Burget, J. Cernock\`y, and S. Khudanpur, 'Recurrent neural network based language model.', in *Interspeech*, 2010, vol. 2, p. 3.
[42] R. Salloum and C.-C. J. Kuo, 'ECG-based biometrics using recurrent neural networks', in *2017 IEEE International Conference on Acoustics, Speech and Signal Processing (ICASSP)*, 2017, pp. 2062–2066.
[43] M. Zihlmann, D. Perekrestenko, and M. Tschannen, 'Convolutional Recurrent Neural Networks for Electrocardiogram Classification', Oct. 2017.
[44] S. Ioffe and C. Szegedy, 'Batch Normalization: Accelerating Deep Network Training by Reducing Internal Covariate Shift', Feb. 2015.
[45] Y. LeCun, Y. Bengio, and G. Hinton, 'Deep learning', *Nature*, vol. 521, no. 7553, pp. 436–444, 2015.
[46] A. L. Goldberger, L. A. Amaral, L. Glass, J. M. Hausdorff, P. C. Ivanov, R. G. Mark, J. E. Mietus, G. B. Moody, C. K. Peng, and H. E. Stanley, 'PhysioBank, PhysioToolkit, and PhysioNet: components of a new research resource for complex physiologic signals.', *Circulation*, vol. 101, no. 23, pp. E215-20, Jun. 2000.
[47] A. ECAR, 'Recommended practice for testing and reporting performance results of ventricular arrhythmia detection algorithms', *Assoc. Adv. Med. Instrum.*, 1987.
[48] D. P. Kingma and J. Ba, 'Adam: A Method for Stochastic Optimization', Dec. 2014.
[49] S. M. Mathews, L. F. Polania, and K. E. Barner, 'Leveraging a discriminative dictionary learning algorithm for single-lead ECG classification', in *2015 41st Annual Northeast Biomedical Engineering Conference (NEBEC)*, 2015, pp. 1–2.